# Discretisation of a stochastic continuum equation of ion-sputtered surfaces

Oluwole Emmanuel Oyewande

*Department of Physics, University of Ibadan, Ibadan, Nigeria.*

*(Dated:16 March 2011)*

## ABSTRACT

*The generalised continuum theory model of the dynamical evolution of surfaces sputtered by ion-bombardment is a noisy Kuramoto-Sivashinsky type partial differential equation. For some generic cases of sputtering parameters, existing similar equations have shed a great deal of light and therefore provided some understanding of the intricacies of evolving ion-sputtered surfaces without a direct solution of the generalised model. However, recent results have demonstrated a wider range of scaling regimes of the sputtering conditions, a large number of which have no similar existing solved models in other research fields for comparison, and whose characteristics are therefore largely unknown. In this paper, a discretisation of the generalised continuum model is performed for direct numerical simulations, the results of which are applicable to all manner of simulations required for the different possible scenarios in the dynamical evolution of sputtered surfaces. The approximation errors and implementation of the results in any such simulation are also discussed.*

## INTRODUCTION

The formation of ordered patterns of nanostructures on material surfaces when the surfaces are bombarded by a beam of ions has been a major focus of research towards advancement in nanotechnology for some time now (Carter & Vishnyakov, 1996; Chason, Mayer, Kellerman, McIlroy, & Howard, 1994; Facsko, Dekorsy, Trappe, Kurz, Vogt, & Hartnagel, 1999; Habenicht, Bolse, Lieb, Reimann, & Geyer, 1999; Carter, 2006; Carter, 2001). Ion bombardment of a material surface sputters the surface atoms and creates a surface instability such that troughs are eroded in preference to crests, as a result of which periodic ripple patterns of nano-dimensions are created on the surface. These ordered nanostructures that are created from the random ejection of surface particles have the interesting characteristic that they orient themselves in either of two directions, depending on the angle of incidence of the impinging ion. For small incidence angles, the ripple structures are oriented perpendicular to the projection of the ion-beam direction onto the surface plane. Whereas for large incidence angles they are oriented parallel to the ion-beam projection (Bradley & Harper, 1988; Barabasi & Stanley, 1995; Habenicht, Bolse, Lieb, Reimann, & Geyer, 1999; Carter & Vishnyakov, 1996). In recent years the materials science community and related fields have witnessed further advances such as the formation of a diverse variety of surface nano-patterns e.g. holes, dots, pits, cones, and so on, all of which depend on the sputtering conditions.

Much understanding of these and other sputtering manifestations like sputtering yield, surface roughness, scaling of interface fluctuations, etc. has been achieved by the continuum theory. This entails a representation of an evolving interface by a continuous and differentiable function $h(x,t)$ and modelling of the dynamics of the interface by stochastic or deterministic partial differential equations. In general, a surface evolving according to the sputtering process is modelled by a stochastic partial differential equation (SPDE) of the form (Cuerno & Barabasi, 1995)

$$\partial_t h(\pmb{x},t) = -v_0 + \zeta \partial_x h(\pmb{x},t) + \varsigma_x \partial_{xx} h(\pmb{x},t) + \varsigma_y \partial_{yy} h(\pmb{x},t) + \eta_x [\partial_x h(\pmb{x},t)]^2 + \eta_y [\partial_y h(\pmb{x},t)]^2 - D\nabla^4 h(\pmb{x},t) + \beta \quad (1)$$

where, as regards ion-bombardment induced sputter erosion, $v_0$ is the erosion velocity of a flat surface, $\zeta$ is a proportionality constant related to the local surface slope along the $x$ –direction, $\varsigma_x$ and $\varsigma_y$ are the (linear) surface tension coefficients, $\eta_x$ and $\eta_y$ are the nonlinear coefficients, $D$ is the (thermal) surface diffusion coefficient, and $\beta$ is a (Gaussian) noise term with zero mean representing the randomness in the ejection of the surface particles.

Up till now Equation 1 has not been solved largely due to the fact that when the model was proposed very few number of scaling regions were found for the small range of sputtering parameters considered. Hence, there was no effort to solve it then as a comparative approach to its solution was taken where the equation, for a particular set of coefficients that represent any of the three regions found, was compared to existing analogous PDEs (e.g. the Kardar-Parisi-Zhang equation for surface growth by deposition) whose solutions were known. Even then a few scenarios had no known solutions (Cuerno & Barabasi, 1995) and up till now their scaling behaviours are largely unclear. Furthermore, to model a surface evolution by Equation 1 there is the need to discretise it for a computer simulation. Complications can arise in this case where errors propagate or where the discretisation is unstable. Moreover, the noise term in Equation 1 makes it non-deterministic and hence non-susceptible to conventional analytical methods. Hence, where the number of different scaling regions are quite small and models exist for comparison the current comparative approach is quicker.

Recent results of phase diagram calculations over a wider range of sputtering parameters have shown the existence and relevance of an even larger number of scaling regions in the phase space defined by the values of the sputtering parameters (Oyewande, 2011). Hence, a numerical solution of Equation 1, subject to appropriate initial conditions, must be performed despite the inherent difficulties. In this paper we discretise Equation 1 and show the condition under which the approximation errors for this discretisation are negligible. We discuss the general implementation of this discrete model in numerical simulations of the sputtered-surface evolution, which can be in a variety of forms depending on the initial conditions to be imposed, the timeframe within which the evolution is to be studied, and the surface characteristics (e.g roughness, structure factor, etc.).

**METHODOLOGY**

Suppose we denote the position of a point $i$ on the surface by $x_i$, then the surface height at this point is $h(x_i)$ and the surface height at an arbitrary position $x = x_i + \Delta x$ is $h(x) = h(x_{i+\Delta x})$, where $\Delta x$ can be positive or negative. The surface heights at the nearest neighbours of $i$ are $h(x_{i+1})$ and $h(x_{i-1})$, for $\Delta x = 1$ and $\Delta x = -1$ respectively. We may write

$$h(x_{i+1}) = h(x_i) + |\Delta x|h'(x_i) + \frac{|\Delta x|^2}{2}h''(x_i) + \frac{|\Delta x|^3}{6}h'''(x_i) + \frac{|\Delta x|^4}{4!}h^{(4)}(x_i)\cdots \quad (2)$$

$$h(x_{i-1}) = h(x_i) - |\Delta x|h'(x_i) + \frac{|\Delta x|^2}{2}h''(x_i) - \frac{|\Delta x|^3}{6}h'''(x_i) + \frac{|\Delta x|^4}{4!}h^{(4)}(x_i)\cdots \quad (3)$$

$$\Rightarrow h(x_{i+1}) - h(x_{i-1}) = 2|\Delta x|h'(x_i) + 2\frac{|\Delta x|^3}{6}h'''(x_i) + 2\frac{|\Delta x|^5}{5!}h^{(5)}(x_i) + \cdots$$

Thus,

$$h'(x_i) = \frac{h(x_{i+1}) - h(x_{i-1})}{2|\Delta x|} - \frac{|\Delta x|^2}{6}h'''(x_i) - \frac{|\Delta x|^4}{5!}h^{(5)}(x_i) - \frac{|\Delta x|^6}{7!}h^{(7)}(x_i)\cdots \quad (4)$$

Now,

$$h(x_{i+1}) + h(x_{i-1}) = 2h(x_i) + |\Delta x|^2 h''(x_i) + 2\frac{|\Delta x|^4}{4!}h^{(4)}(x_i) + 2\frac{|\Delta x|^6}{6!}h^{(6)}(x_i) + \cdots$$

$$\Rightarrow h''(x_i) = \frac{h(x_{i+1}) - 2h(x_i) + h(x_{i-1})}{|\Delta x|^2} - \frac{|\Delta x|^2}{12}h^{(4)}(x_i)\cdots \quad (5)$$

$$h^{(4)}(x_i) = (4!)\frac{h(x_{i+1}) - 2h(x_i) - |\Delta x|^2 h''(x_i) + h(x_{i-1})}{2|\Delta x|^4} - \frac{|\Delta x|^2}{30}h^{(6)}(x_i) - \cdots \quad (6)$$

The foregoing provides the discrete forms of the terms in Equation 1, and is applied to discretise the equation. The results are provided in the next section.

**RESULTS AND DISCUSSION**

Using Equations 4 and 5 the central difference approximations in space to the partial derivatives of the continuum equation are

$$[h_x(x,y)]_{x_i} = \frac{h(x_{i+1},y) - h(x_{i-1},y)}{2|\Delta x|} \quad (7)$$

$$[h_y(x,y)]_{y_i} = \frac{h(x,y_{i+1}) - h(x,y_{i-1})}{2|\Delta y|} \quad (8)$$

with an approximate error or first difference of $\frac{|\Delta x|^2}{6}h'''(x_i)$,

$$[h_{xx}(x,y)]_{x_i} = \frac{h(x_{i+1},y) - 2h(x_i,y) + h(x_{i-1},y)}{|\Delta x|^2} \quad (9)$$

$$[h_{yy}(x,y)]_{y_i} = \frac{h(x,y_{i+1}) - 2h(x,y_i) + h(x,y_{i-1})}{|\Delta y|^2} \quad (10)$$

with an approximate error of $\frac{|\Delta x|^2}{12}h^{(4)}(x_i)$. Equation 6 leads to

$$\left[\frac{\partial^4 h(x,y)}{\partial x^4}\right]_{x_i} = (4!)\frac{h(x_{i+1},y) - 2h(x_i,y) - [h(x_{i+1},y) - 2h(x_i,y) + h(x_{i-1},y)] + h(x_{i-1},y)}{2|\Delta x|^4}$$

$$\Rightarrow \left[\frac{\partial^4 h(x,y)}{\partial x^4}\right]_{x_i} = \left[\frac{\partial^4 h(x,y)}{\partial y^4}\right]_{y_i} = 0 \quad (11)$$

$$\frac{\partial^4 h(x,y)}{\partial y^2 \partial x^2} = \left(h_{yy}[h_{xx}(x,y)]_{x_i}\right)_{y_i} = \frac{h(x_{i+1},y_{i+1}) - 2h(x_i,y_i) + h(x_{i-1},y_{i-1})}{|\Delta y|^2 |\Delta x|^2}$$

$$= \frac{\partial^4 h(x,y)}{\partial x^2 \partial y^2} \quad (12)$$

with an approximate error of $\frac{|\Delta x||\Delta y|}{30} h^{(6)}(x_i)$.

Using forward difference in time we have

$$[h_t(x,y,t)]_{t_i} = \frac{h(x,y,t_{i+1}) - h(x,y,t_i)}{|\Delta t|} \quad (13)$$

with an approximate error of $\frac{|\Delta t|}{2} h''(t)$.

The surface diffusion term is given by

$$\nabla^4 h = \nabla \cdot \nabla \cdot \nabla \cdot \nabla \cdot h = \nabla \cdot \nabla \cdot \nabla \cdot \left(i\frac{\partial h}{\partial x} + j\frac{\partial h}{\partial y}\right)$$

$$\Rightarrow \nabla^4 h = \frac{\partial^4 h}{\partial x^4} + \frac{\partial^4 h}{\partial x^2 \partial y^2} + \frac{\partial^4 h}{\partial y^2 \partial x^2} + \frac{\partial^4 h}{\partial y^4}$$

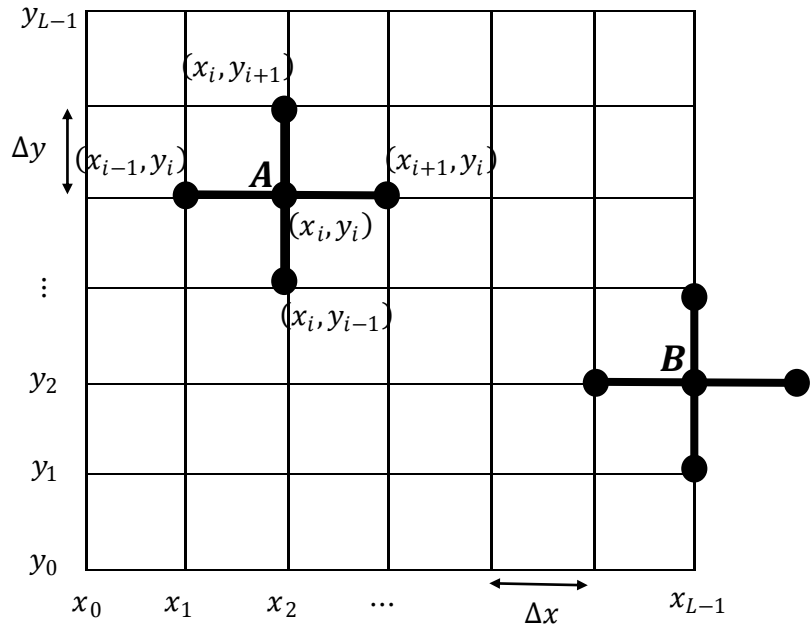

Figure 1: Discrete representation of a surface as a square lattice of linear size *L*. Point *A* lies within the lattice and so does its four nearest neighbours. Point *B* lies on the lattice boundary and one of its nearest neighbours is off the lattice. The dots (nodes) are lattice sites representing surface points, and the bold lines connecting them (edges) represent interactions between the points.

According to the result (11) which is based on our central difference approximation for $h''(x_i)$,

$$\nabla^4 h \approx \frac{\partial^4 h}{\partial x^2 \partial y^2} + \frac{\partial^4 h}{\partial y^2 \partial x^2} \quad (14)$$

$$\Rightarrow \nabla^4 h \approx \frac{2[h(x_{i+1}, y_{i+1}) - 2h(x_i, y_i) + h(x_{i-1}, y_{i-1})]}{|\Delta y|^2 |\Delta x|^2} \quad (15)$$

which concludes the discretisation. The result of the analysis of approximation errors indicate that they are negligible for $\Delta x \ll 1, \Delta y \ll 1, \Delta t \ll \Delta x$.

A numerical simulation of the time evolution of a sputtered surface governed by Equation 1 may now be performed by using the discretised version of the stochastic continuum equation on a mesh of appropriately chosen grid size as illustrated in Fig. 1. A square lattice may be used for this purpose of simulating the surface. Points that are at the boundary are different to those within the lattice in having a fewer number of neighbours (the rest being off the lattice), as in Fig. 1. This will affect the results and may be corrected by imposing periodic boundary conditions such that the point on the boundary is regarded as the corresponding point on the other end of the lattice; with the effect that the planar square lattice may be seen as a torus. An implementation of the periodic boundary conditions may therefore be carried out by setting:

$$\begin{aligned} h(x_{L-1}, y_{L-1}) &= h(x_0, y_0) \\ h(x_{L-1}, y_i) &= h(x_0, y_i) \\ h(x_i, y_{L-1}) &= h(x_i, y_0) \end{aligned} \quad (16)$$

**CONCLUSION AND RECOMMENDATIONS**

While the dynamical evolution of sputtered surfaces are readily modelled and analysed with a generalised stochastic continuum equation, very few of the scaling regions predicted by this continuum model are understood due to the very limited knowledge of its numerical solution, complicated by the complexity of the sputtering process. In this paper, a generalised approach to its numerical solution for diverse sputtering conditions has been presented. The crucial part of this approach is the discretisation of the stochastic continuum equation such that the discrete model derived can be simulated on a mesh, subject to appropriate initial conditions (e.g. an initially flat surface, an initially disordered surface, a ripple surface etc.). Through such a simulation, analysis of various characteristics of the evolving surface can be performed and knowledge of the scaling behaviour and other characteristics of the SPDE can be achieved.

Details of the discretisation have been presented in this work along with estimates of the approximation errors, which are negligible for small lattice separations and step size. A discussion on how to implement the simulation has also been provided. It is recommended that simulations be performed for different sputtering scenarios. Also, since SPDEs are widely applicable in different fields, these results will have applications extending beyond the present scope and such are recommended. Finally, while the central difference approximations applied here are the most efficient and stable, stability analysis could still prove useful and may lead to a more efficient algorithm.

e-mail: eoyewande@gmail.com